\documentclass[english]{article}
\usepackage{mathptmx}
\usepackage{helvet}
\usepackage{courier}
\usepackage[T1]{fontenc}
\usepackage[latin9]{inputenc}
\usepackage[a4paper]{geometry}
\usepackage{float}
\usepackage{setspace}
\makeatletter

\providecommand{\tabularnewline}{\\}

\makeatother

\usepackage{babel}

\begin{document}

\title{Comment on {}``Semiquantum secret sharing using entangled states''}

\date{\,}

\begin{doublespace}

\author{Jason Lin, Chun-Wei Yang, Chia-Wei Tsai, and Tzonelih Hwang%
\thanks{Corresponding Author%
}}
\end{doublespace}
\maketitle
\begin{abstract}
Recently, Li et al. {[}Phys. Rev. A, 82(2), 022303{]} presented two
semi-quantum secret sharing (SQSS) protocols using GHZ-like states.
The proposed schemes are rather practical because only the secret
dealer requires to equip with advanced quantum devices such as quantum
memory, whereas the other agents can merely perform classical operations
to complete the secret sharing. However, this study points out that
a security pitfall exists in the eavesdropping check phase of both
schemes that could mount to an Intercept-resend attack and a Trojan
horse attack on the two schemes, respectively, to disclose the other
agent's shadow, and further to reveal the master key of the SQSS,
which contradicts to the security requirement of a QSS. Fortunately,
two possible solutions are proposed to avoid this security pitfall.

\textbf{\emph{keywords:}} Quantum secret sharing, GHZ-like state,
Intercept-resend attack, Trojan horse attack
\end{abstract}

\section{Introduction}

Since the first quantum secret sharing (QSS) protocol was presented
by Mark et al.'s via triplet Greenberger-Horne-Zeilinger (GHZ) state
in 1999 \cite{QSS-HBB}, lots of QSS schemes have also been proposed
{[}2-13{]}. The main goal of a QSS is to distribute a secret among
several agents based on the quantum mechanics. Only when enough subsets
of legitimate agents cooperate can the secret be recovered. On the
contrary, any agent alone is not able to acquire the dealer's secret
by his/her own shadow. A secure QSS should be able to avoid the attack
from both an outside eavesdropper and an inside malicious user.

Recently, Li et al. proposed two novel semi-quantum secret sharing
(SQSS) protocols via triplet GHZ-like state \cite{QSS-Li}. According
to their definition, the term {}``semi-quantum'' implies that the
secret dealer is a powerful quantum server, whereas the other agents
are all classical clients. More precisely, the secret dealer has the
ability to perform the following operations: (1) preparing GHZ-like
state, (2) performing the Bell measurement and the three-qubit joint
measurement, (3) storing photons in a short-term quantum memory. As
for the classical agents, they are restricted to perform the following
operations over the quantum channel: (1) preparing new qubits in the
classical basis \{$\left|0\right\rangle $, $\left|1\right\rangle $\},
(2) measuring photons in the classical basis, (3) reordering the photons
via different delay lines, (4) sending or reflecting the qubits without
disturbance. Since the classical basis only considers the qubit $\left|0\right\rangle $
and $\left|1\right\rangle $, the other quantum superpositions of
single photon are not included here. Therefore, the agents' operations
above are equivalent to the traditional \{0, 1\} computation.

The two protocols proposed by Li et al. \cite{QSS-Li} are namely
the randomization-based SQSS and the measure-resend SQSS, respectively.
Both schemes are based on the entanglement correlation of GHZ-like
state $\left|\psi'\right\rangle =\frac{1}{2}\left(\left|000\right\rangle +\left|011\right\rangle +\left|110\right\rangle +\left|101\right\rangle \right)=\frac{1}{\sqrt{2}}\left(\left|0\right\rangle \left|\phi^{+}\right\rangle +\left|1\right\rangle \left|\psi^{+}\right\rangle \right)$,
which can be easily generated by performing the Hadamard gate $H$
($=\frac{1}{\sqrt{2}}\left(\left|0\right\rangle \left\langle 0\right|+\left|0\right\rangle \left\langle 1\right|+\right.$
$\left.\left|1\right\rangle \left\langle 0\right|-\left|1\right\rangle \left\langle 1\right|\right)$)
on each qubit of the standard GHZ state $\left|\Psi_{1}\right\rangle =\frac{1}{\sqrt{2}}\left(\left|000\right\rangle +\left|111\right\rangle \right)$.
Under the three-party QSS scenario, it can be seen that if each party
holds the \textit{1st},\textit{ }the\textit{ 2nd}, and the \textit{3rd}
particle of a GHZ-like state, respectively, then their classical-basis
measurements (say $MR_{1}$, $MR_{2}$, and $MR_{3}$) will agree
to a secret sharing relationship: $MR_{1}=MR_{2}\oplus MR_{3}$, where
the measurement result is encoded as '0' if $\left|0\right\rangle $,
'1' if $\left|1\right\rangle $.

However, this study attempts to show that under the three-party scenario
(i.e., one boss and two agents) of Li et al.'s scheme, a malicious
agent is possible to launch an Intercept-resend attack on the randomization-based
SQSS and a Trojan horse attack \cite{TH-attack-1,TH-attack-2,TH-attack-3,TH-attack-4}
on the measure-resend SQSS to reveal the other agent's shadow. This
contradicts to the security requirements of a QSS. Fortunately, the
above problems can be respectively solved by a carefully designed
eavesdropping check process and the use of some special optical devices
that filter out the spy photons of the Trojan horse attacks.

The rest of this paper is constructed as follows. Section 2 reviews
Li et al.'s two SQSS schemes via GHZ-like state. Section 3 points
out the problem and gives two solutions to remedy the loophole. Finally,
Section 4 gives a brief conclusion to the result.

\section{Review of Li et al.'s SQSS schemes}

In this section, a brief review of Li et al.'s two SQSS schemes is
given. The only difference between these two schemes is the definition
of the classical agent's ability. For a randomization-based SQSS protocol,
classical agents are limited to perform operations: (2), (3), and
(4), while in a measure-resend protocol, classical agents are limited
to perform operations: (1), (2), and (4), as defined in Sec. 1.

\subsection{Randomization-based SQSS protocol}

In this subsection, the SQSS is considered under a three-party scenario
as follows. Suppose a boss Alice wants to share a secret with her
two agents: Bob and Charlie. She splits her secret key $K_{A}$ into
two pieces of shadow key: $K_{B}$ and $K_{C}$, which will deliver
to Bob and Charlie, respectively. Only when Bob and Charlie collaborate
can $K_{A}$ be recovered. The procedure of the randomization-based
SQSS can be described in the following steps:
\begin{description}
\item [{Step~1.}] Alice first prepares $N$ triplet GHZ-like states all
in $\left|\psi'\right\rangle =\frac{1}{2}\left(\left|000\right\rangle +\left|011\right\rangle +\left|110\right\rangle +\left|101\right\rangle \right)$.
Here, the quantum states $\left\{ \left|0\right\rangle ,\left|1\right\rangle \right\} $
can be classically measured by $Z$ basis. Then, she divides these
$N$ GHZ-like states into three sequences $S_{A}$, $S_{B}$, and
$S_{C}$, which include the \textit{1st}, the \textit{2nd}, and the
\textit{3rd} particles of all GHZ-like states, respectively. After
the above preparation, Alice retains the quantum sequence $S_{A}$,
and sends the sequence $S_{B}$ to Bob, $S_{C}$ to Charlie.
\item [{Step~2.}] When Bob and Charlie receive the photons, respectively,
they choose to adopt either the SHARE mode or the CHECK mode on each
qubit, respectively. In the SHARE mode, the agent performs a $Z$-basis
measurement on the qubit, whereas in the CHECK mode, the agent reflects
the qubit back to Alice. Notice that those returned qubits in the
CHECK mode are reordered via different delay lines.
\item [{Step~3.}] Alice stores the reflected qubits from Bob and Charlie
in a short-term quantum memory, and publicly announces the reception
of these photon sequences. After that, Bob and Charlie publish the
correct order of the reflected qubits, and their original positions
in the sequences delivered by Alice, respectively. According to the
agents' reports, Alice can recover the reflected qubits into the correct
order.
\item [{Step~4.}] For each GHZ-like state, both Bob and Charlie announce
their decisions respectively on the corresponding two particles of
$S_{B}$ and $S_{C}$, which can be one of the four cases as shown
in Table 1. Then, Alice can perform one of the four actions on the
corresponding qubits as depicted in Table 1.
\item [{Step~5.}] For the eavesdropping check, those qubits in cases (2),
(3), and (4) of Table 1 are publicly discussed. The involved parties
have to publish their measurement results in those cases to see whether
each corresponding three qubits is consistent to the correlation of
a GHZ-like state $\left|\psi'\right\rangle $ ($=\frac{1}{2}\left(\left|000\right\rangle +\left|011\right\rangle +\left|110\right\rangle +\left|101\right\rangle \right)=\frac{1}{\sqrt{2}}\left(\left|0\right\rangle \left|\phi^{+}\right\rangle +\left|1\right\rangle \left|\psi^{+}\right\rangle \right)$).
If the error rate is higher than a predetermined threshold, then Alice
terminates the protocol and restarts from Step 1. Otherwise, the protocol
continues to the next step.
\item [{Step~6.}] As for the secret sharing policy in the case (1) of
Table 1, the \textit{1st},\textit{ }the\textit{ 2nd}, and the \textit{3rd}
qubits of GHZ-like states are measured by Alice, Bob, and Charlie,
respectively, using $Z$-basis. They can transform these measurement
results into three binary bit sequences, in which the result is '0'
if $\left|0\right\rangle $ and '1' if $\left|1\right\rangle $. After
the transformation, Alice, Bob, and Charlie will obtain a key bit
string $K_{A}$, $K_{B}$, and $K_{C}$, respectively, which conform
to the secret sharing relationship, i.e., $K_{A}=K_{B}\oplus K_{C}$.
\end{description}
\begin{table}[H]
\caption{The actions taken by the secret dealer Alice in each case.}

$\vphantom{}$

\begin{centering}
\begin{tabular}{c|c|c|c}
\hline 
Case & Bob & Charlie & Alice\tabularnewline
\hline
\hline 
(1) & SHARE & SHARE & ACTION (i)\tabularnewline
\hline 
(2) & SHARE & CHECK & ACTION (ii)\tabularnewline
\hline 
(3) & CHECK & SHARE & ACTION (iii)\tabularnewline
\hline 
(4) & CHECK & CHECK & ACTION (iv)\tabularnewline
\hline
\end{tabular}
\par\end{centering}

$\vphantom{}$
\begin{description}
\item [{{\small (i):}}] {\small Alice measures her own qubit with $Z$-basis.}{\small \par}
\item [{{\small (ii):}}] {\small Alice performs Bell measurement on her
qubit and Charlie's returned qubit.}{\small \par}
\item [{{\small (iii):}}] {\small Alice performs Bell measurement on her
qubit and Bob's returned qubit.}{\small \par}
\item [{{\small (iv):}}] {\small Alice performs an appropriate three-qubit
joint measurement on her qubit and the returned qubits.}
\end{description}

\end{table}

The randomization-based SQSS protocol uses the entanglement correlation
of GHZ-like state $\left|\psi'\right\rangle $ to achieve the goal
of secret sharing. In this type of protocol, the agents will directly
perform $Z$-basis measurement on the photons in the SHARE mode. Conversely,
by modifying the operations performed by the agents, Li et al. further
proposed the other scheme called the measure-resend SQSS protocol,
which will be described in Sec. 2.2.

\subsection{Measure-resend SQSS protocol}

Similar to Sec. 2.1, the measure-resend SQSS scheme is also reviewed
under a three-party scenario (i.e., a boss Alice, and two agents:
Bob and Charlie). The modified steps ({*}) are depicted in detail
as follows. The other steps are the same as those described in Sec.
2.1 and thus are omitted here.
\begin{description}
\item [{({*}Step~2)}] There are two modes (i.e., SHARE and CHECK) that
Bob and Charlie can decide to perform on each received photon. For
the CHECK mode, the agent still reflects the qubit back to Alice via
different delay lines similar to Sec. 1. On the contrary, in the SHARE
mode, the agent measures the received qubits in $Z$-basis, and returns
a sequence of newly generated photons of the same states to Alice.
\item [{({*}Step~3)}] Alice stores the photon sequences reflected from
Bob and Charlie in a short-term quantum memory, and publicly confirms
the reception of them. Subsequently, Bob and Charlie declare the positions
of particles being measured and being reflected.
\item [{({*}Step~4)}] According to the agents' reports, Alice can perform
one of the four actions on her own qubit and the corresponding qubits
as depicted in Table 1.
\end{description}
The measure-resend SQSS protocol is also based on the entanglement
correlation of the GHZ-like state $\left|\psi'\right\rangle $. The
only difference between these two schemes (the randomization-based
SQSS and the measure-resend SQSS) is the type of operations allowed
to perform by the agent in the SHARE mode. Considering the eavesdropping
check, both schemes discuss the measurement result of each qubit in
the GHZ-like state to detect the presence of eavesdroppers. However,
this check strategy may not be able to prevent Bob or Charlie from
maliciously launching attacks on the SQSS protocols. More details
of the attacks will be discussed in Sec. 3.

\section{Attacks and the improvements}

This section shows that under the three-party scenario (i.e., one
boss and two agents) of Li et al.'s scheme, a malicious agent is possible
to launch an Intercept-resend attack on the randomization-based SQSS
and a Trojan horse attack \cite{TH-attack-1,TH-attack-2,TH-attack-3,TH-attack-4}
on the measure-resend SQSS to reveal the other agent's shadow and
further to derive Alice's secret key. This contradicts to the security
requirements of a QSS. Fortunately, the above problems can be respectively
solved by a carefully designed eavesdropping check process and the
use of some special optical devices that filter out the spy photons
of the Trojan horse attacks.

\subsection{Attacks on Li et al.'s SQSS schemes}

Both Bob and Charlie can act as a dishonest insider to derive Alice's
shared secret. In general, an eavesdropper is assumed to be powerful
enough to equip with any quantum devices \cite{Eve-1,Eve-2,Eve-3}.
Hence, the malicious classical agent is able to perform any operation
as defined in Sec. 1.

\subsubsection{The Intercept-resend attack on the randomization-based SQSS.}

Suppose that Bob is a dishonest insider. He first intercepts the photon
sequence $S_{C}$ (from Alice to Charlie) in Step 1, and stores it
in his quantum memory. Then, he prepares a new photon sequence $S_{E}$
randomly chosen from $\left|0\right\rangle $ or $\left|1\right\rangle $,
and sends it to Charlie, where $S_{E}$ is of the same length as $S_{C}$.
Notice that the wavelength of each photon in $S_{E}$ is set to be
different from the others so that Bob is alble to identify their individual
position.

When Charlie receives the sequence $S_{E}$ in Step 2, he will perform
$Z$-basis measurement on those photons chosen for the SHARE mode,
and reflect the ones that are chosen for the CHECK mode via different
delay lines. At this time, Bob can intercept the reflected sequence
(from Charlie to Alice), and replace those photons with the corresponding
photons in $S_{C}$ and then send them back to Alice. Bob is able
to do so by distinguishing the wavelengths of the reflected photons
from Charlie. 

Later, Bob deliberately selects the SHARE mode on those photons in
$S_{B}$ that their corresponding photons in $S_{C}$ have been chosen
by Charlie as in the SHARE mode, and randomly select SHARE or CHECK
on the other photons in $S_{B}$. The above action is to avoid the
presence of the case (3) in Table 1 because it has a $50\%$ probability
of being detected. More precisely, since all the SHARE photons measured
by Charlie are the forged photons in $S_{E}$, there is a $50\%$
probability on each three-particle set of the case (3) that will not
follow the entanglement correlation of GHZ-like state $\left|\psi'\right\rangle =\frac{1}{\sqrt{2}}\left(\left|0\right\rangle \left|\phi^{+}\right\rangle +\left|1\right\rangle \left|\psi^{+}\right\rangle \right)$.

For the eavesdropping check, Bob can escape from detection because
of all the reflected photons in cases (1), (2), and (4) of Table 1
are indeed generated by Alice. Therefore, he can obtain Charlie's
shadow $K_{C}$ by measuring the SHARE photons in $S_{C}$, and further
derive Alice's secret key with $K_{B}\oplus K_{C}=K_{A}$.

\subsubsection{The Trojan-horse attack on the measure-resend SQSS.}

Let us also assume here that Bob is a malicious insider. He first
attaches some invisible photons \cite{TH-attack-2,TH-attack-4} $S_{T}$
on each particle of $S_{C}$ transmitted from Alice to Charlie in
Step 1, and then inserts some delay photons \cite{TH-attack-2,TH-attack-3}
$S_{D}$ in the same time window to each particle of $S_{C}$. Notice
that the wavelength in each photon of $S_{D}$ is set to be the same
as the corresponding photon in $S_{C}$, whereas the wavelength in
each photon of $S_{T}$ is close to the corresponding photon in $S_{C}$.

When Charlie receives the sequence $S_{C}$ in Step 2, he measures
those photons in the SHARE mode with $Z$-basis, and returns a sequence
of newly generated photons of the same states to Alice. The corresponding
photons of the SHARE photons in $S_{T}$ and $S_{D}$ will vanish
after the replacement of the newly produced photons. As for the CHECK
photons, Charlie will directly reflect them without any reordering
operation to Alice. At this time, Bob can intercept the returned sequence
(from Charlie to Alice), and perform $Z$-basis measurement on those
photons that their corresponding spy photons have disappeared.

After the measurement, Bob resends the returned sequence back to Alice
without any further action. Since Alice will also perform $Z$-basis
measurement on the SHARE photons of Charlie in Step 4, the measurement
results will not be different from the ones measured by Bob. Hence,
the three cases (1), (2), and (3) in Table 1 used for the eavesdropping
check will not detect the attack. Bob can obtain Charlie's shadow
$K_{C}$ by those $Z$-basis measurement results of the SHARE photons
in the case (4) of Table 1 and further derive Alice's secret key with
$K_{B}\oplus K_{C}=K_{A}$.

\subsection{Possible solutions for the attacks}

Two solutions to avoid the attacks are proposed here. The first one
is to set a new threshold of eavesdropping check in the randomization-based
SQSS. The second solution is to equip with some special optical filter
devices to detect the Trojan horse attacks on the measure-resend SQSS.
\begin{description}
\item [{Solution~1.}] A new threshold for the eavesdropping check.
\end{description}
In Table 1, all four cases should be evenly distributed. However,
if Bob performs the intercept-resend attack as shown in Sec. 3.1.1,
there is no chance for case (3) of Table 1 to appear. Thus, to prevent
this attack, before the eavesdropping check of Step 5, Alice can first
calculate the occurrence $\rho$ of case (3) in Table 1, and decide
the existence of the attack. If $\rho$ is too small, then Alice can
abort the protocol.
\begin{description}
\item [{Solution~2.}] Agents install some optical filter devices.
\end{description}
Since the attack in Sec. 3.1.2 is based on the spy photons in the
Trojan horse attacks, when Charlie receives the photons in Step 2,
he can equip with some special optical devices such as the wavelength
quantum filter and the photon number splitters (PNS) to detect the
attacks. According to \cite{TH-attack-1,TH-attack-2,TH-attack-3,TH-attack-4},
the wavelength quantum filter can eliminate the invisible photons
attached on the legitimate ones, and the PNS can spit each legitimate
particle to discover the delay photons. If there is an irrational
high rate of multi-photon signal, then Charlie announces to restart
the protocol from Step 1.

\section{Conclusions}

This paper has pointed out two attacks on both of Li et al.'s SQSS
schemes, respectively. Under the three-party scenario (i.e., one boss
and two agents), a malicious insider could possibly launch the Intercept-resend
attack on the randomization-based SQSS and the Trojan horse attacks
on the measure-resend SQSS to obtain the other agent's shadow, which
can also lead to derive the boss's secret key. Fortunately, two solutions
are given in this paper to avoid the attacks (i.e., one is to add
a new threshold for the eavesdropping check, and the other is to equip
with some special optical devices to filter out the spy photons).
With the second solution, since near a half of the transmitted photons
are used in devices to detect the Trojan horse attack for each agent,
the qubit efficiency will be seriously jeopardized. Hence, how to
design a QSS protocol which is congenitally free from this attack
is a promising future research.

\section*{Acknowledgement}

We would like to thank the National Science Council of the Republic
of China, Taiwan for partially supporting this research in finance
under the Contract No. NSC 98-2221- E-006-097-MY3.

\end{document}